\documentclass[prl,letterpaper,twocolumn,showpacs]{revtex4}

\usepackage{graphicx}

\newcommand{\be}{\begin{equation}} 
\newcommand{\ee}{\end{equation}}
\newcommand{\bea}{\begin{eqnarray}} 
\newcommand{\eea}{\end{eqnarray}}

\newcommand{\tab}{Table~\ref}

\newcommand{\eqn}{Eq.~\ref}

\newlength{\diaght}
\newlength{\diagshift}
\newcommand{\mayeriva}{\raisebox{\diagshift}{\includegraphics[height=\diaght]{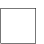}}}
\newcommand{\mayerivc}{\raisebox{\diagshift}{\includegraphics[height=\diaght]{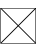}}}
\newcommand{\insivb}{\raisebox{\diagshift}{\includegraphics[height=\diaght]{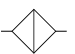}}}
\newcommand{\wigglya}{\ensuremath{\emptyset}}
\newcommand{\wigglyb}{\raisebox{\diagshift}{\includegraphics[height=\diaght]{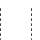}}}

\begin{document}

\title{New results for the virial coefficients of $D$--dimensional hard spheres}

\author{N.~Clisby}
\email{Nathan.Clisby@stonybrook.edu}

\author{B.~M.~McCoy}
\email{mccoy@insti.physics.sunysb.edu}

\affiliation{C.~N.~Yang Institute for Theoretical Physics \\ State
University of New York at Stony Brook \\ Stony Brook, NY 11794-3840}

\date{\today}

\begin{abstract}
Exact results are given for the fourth virial coefficient of hard
spheres in even dimensions up through 12. The fifth and sixth virial
coefficients are numerically computed for dimensions 2 through 50 and
it is found that the sixth virial coefficient is negative for $D\geq
6.$ Numerical studies are made of the contributing Ree Hoover diagrams
up to order 17. It is found for $D\geq 3$ that for large order a class
of diagrams we call ``loose packed'' dominates and the rate of growth
of these diagrams is used to study bounds on the radius of
convergence.
\end{abstract}


\maketitle

The virial series for the pressure \be \frac{P}{k_BT}=
\rho+\sum_{k=2}^{\infty}B_k \rho^{k},
\label{vir}
\ee for hard particles of diameter $\sigma$ where the pair potential
$U({\bf r})$ is $+\infty$ for $|{\bf r}|<\sigma$ and $0$ otherwise,
has been studied in dimensions two and three for over 100
years. However, despite the long history of this problem, the only
rigorous information of the radius of convergence comes from the upper
bound of Lebowitz and Penrose~\cite{lebowitz1964a} that $|{B_k/
B_2^{k-1}}| \leq 13.8^{k-1}/k$ which in terms of the packing fraction
$\eta={B_2\rho / 2^{D-1}}$ gives convergence for $|\eta|\leq {0.145/
2^{D}}$. For $D=3$ this gives a lower bound of $0.018$ which is much
less than the packing fraction at which freezing occurs of
$\eta_f=0.49$~\cite{alder1957a,hoover1968a}.

Our numerical knowledge of the virial expansion for $D=2,3$ is
compactly summarized in \tab{virialtable}. Virial coefficients have
also been calculated for hard spheres in dimensions higher than three
in~\cite{ree1964b,loeser1993a,luban1982a,bishop1999a}.
\begin{table}[hb]
\caption{$B_k/B_2^{k-1}$ for $k=3,\ldots,8$ in $D=2,3$.}
\label{virialtable}
\begin{tabular}{lll}\hline \hline
&discs&spheres\\ \hline $B_2$&$\pi \sigma^2/2$&$2\pi \sigma^3/3$ \\
$B_3/B_2^2$&${4\over 3}-{{\sqrt 3}\over
\pi}$~\cite{tonks1936a}&${5\over 8}$~\cite{boltzmann1899a}\\
$B_4/B_2^3$&$0.5322318\cdots$~\cite{rowlinson1964a,hemmer1964a}
&$0.2869495\cdots$~\cite{boltzmann1899a,vanlaar1899a,nijboer1952a}\\
$B_5/B_2^4$&$0.33355604(4)$~\cite{ree1964a,kratky1982a}
&$0.110252(1)$~\cite{ree1964a,kratky1977a}\\
$B_6/B_2^5$&$0.19883(1)$~\cite{ree1964a,kratky1982c}
&$0.038808(55)$~\cite{ree1964a,vanrensburg1993a}\\
$B_7/B_2^6$&$0.114877(11)$~\cite{ree1967a,vanrensburg1993a,vlasov2002a}
&$0.013046(22)$~\cite{ree1967a,vanrensburg1993a,vlasov2002a}\\
$B_8/B_2^7$&$0.065030(31)$~\cite{vanrensburg1993a,vlasov2002a}&$0.004164(16)$~\cite{vanrensburg1993a,vlasov2002a}\\
\hline \hline
\end{tabular}
\end{table}

There have been many attempts to estimate a radius of convergence and
to find an approximate low density form for the equation of state
based on these first few virial coefficients. In three dimensions
these approximates may be grouped into three classes in terms of the
location of the leading singularity in the $\eta$ plane. Either there
is (1) a high order
pole~\cite{thiele1963a,wertheim1963a,guggenheim1965a,carnahan1969a} at
$\eta=1$, (2) a simple
pole~\cite{hoover1968a,ree1964a,vanrensburg1993a,hoste1984a,goldman1988a}
at or near the packing fraction $\eta_{cp}=0.74048\cdots$ of closest
packed spheres, or (3) a fractional power law
divergence~\cite{lefevre1972a,ma1986a,jasty1987a,song1988a} at or near
the ``random close packed'' density $\eta_{rcp}=0.64$ as defined
by~\cite{bernal1964a}.

All of these approximates have radii of convergence greater than the
endpoints of the numerically determined~\cite{alder1957a,hoover1968a}
first order phase transition $\eta_f=0.49$ and $\eta_s=0.54$; because
of this it is often assumed that the virial expansion for hard spheres
is analytic at the freezing density $\eta_f$. This analyticity
assumption is incorporated in most of the phenomenological theories
used to describe
freezing~\cite{kirkwood1940a,ramakrishnan1979a,haymet1986a}.
There is clearly a large discrepancy between the lower bound on the
radius of convergence and the assumption of analyticity at $\eta_f$.

Furthermore all the known virial coefficients for $D=2,3,4,5$ are
positive and this positivity is often built into the approximate
equations of state by having the leading singularity on the positive
real axis. But the possibility of negative virial coefficients was
suggested by Temperley~\cite{temperley1957a} as far back as 1957, and
in 1964 it was shown~\cite{ree1964b} that $B_4$ is negative for $D\geq
8$.  The sixth and seventh virial coefficients are negative for
parallel hard cubes~\cite{hoover1962a} and oscillatory signs are found
in models of hard squares~\cite{baxter1970a} and hard
hexagons~\cite{baxter1980a,joyce1988a}.  Consequently there is no {\it
a priori} reason to expect the leading singularity to be on the real
axis.

We have studied these questions of sign change and rate of growth of
the virial coefficients by numerically, and for $B_4$ analytically,
evaluating selected Ree Hoover diagrams for orders up through 17.  We
use the definitions and notations adapted from~\cite{ree1964c}. In
particular, for diagrams of $k$ points which contribute to $B_k$, each
point is connected to each other point either by an $f$ bond that
corresponds to the function $f({\bf r})$ which is $-1$ for $|{\bf r}|<
\sigma$ and $0$ otherwise, or by a bond ${\tilde f}=1+f.$ In drawing
the diagrams we need only specify either $f$ bonds (represented by
solid lines) or ${\tilde f}$ bonds, not both. For example the fourth
virial coefficient is given by \be B_4 =\frac{1}{4}\, \wigglya -
\frac{3}{8} \, \wigglyb =\frac{1}{4}\, \mayerivc - \frac{3}{8} \,
\mayeriva.
\label{b4rheq}
\ee The first expression in \eqn{b4rheq} is the expansion in
Ree-Hoover graphs with $\tilde f$ bonds, while the second is the same
expansion but with only $f$ bonds shown.

The integrals in \eqn{b4rheq} have been previously evaluated
analytically only for $D=2$~\cite{rowlinson1964a,hemmer1964a} and
$3$~\cite{boltzmann1899a,vanlaar1899a,nijboer1952a}. We have extended
these analytic computations for even dimensions up through $D=12.$ The
results are shown in \tab{analyticalB4table}. Details of the
computation will be published elsewhere.
\begin{table}[thb]
\caption{Analytical results for $B_4/B_2^3$ in even dimensions.}
\label{analyticalB4table}
\begin{tabular}{llr} 
\hline \hline $D$ & Analytic Value & Numerical Value \\ \hline 2 &
$2-\frac{9\sqrt{3}}{2\pi}+\frac{10}{\pi^2}$&$0.53223180\cdots$\\ 4 &
$2-\frac{27\sqrt{3}}{4\pi}+\frac{832}{45\pi^2}$ &$0.15184606\cdots$\\
6 & $2-\frac{81\sqrt{3}}{10\pi}+\frac{38848}{1575\pi^2}$&$
0.03336314\cdots$\\ 8 &
$2-\frac{2511\sqrt{3}}{280\pi}+\frac{17605024}{606375\pi^2}$
&$-0.00255768\cdots$\\ 10 &
$2-\frac{2673\sqrt{3}}{280\pi}+\frac{49048616}{1528065\pi^2}$
&$-0.01096248\cdots$ \\ 12 &
$2-\frac{2187\sqrt{3}}{220\pi}+\frac{11565604768}{337702365\pi^2}$
&$-0.01067028\cdots$ \\ \hline \hline
\end{tabular}
\end{table}

To investigate the phenomena of negative virial coefficients further
we have made a Monte-Carlo evaluation of $B_4, B_5,$ and $B_6$ for
dimensions up to 50. The method used allows the calculation of $B_k$
for dimensions $D \ge k-1$, including non-integer dimensions, and will
be reported elsewhere. The results for integer dimensions up to 12 are
shown in \tab{B456table} where the entries for $B_6$ in $D=4$ and $5$
are from~\cite{bishop1999a}. For dimensions higher than 12 these
$B_k/B_2^{k-1}$ approach zero in a monotonic fashion.  From an
interpolation of these results we see that $B_4$ becomes negative at
$D = 7.73$, $B_6$ becomes negative at $D\sim5.3,$ and that while
$B_5/B_2^4$ is always positive it is not monotonic. The fact that the
zero crossing of $B_k$ has decreased from $7.73$ to $5.3$ as $D$
increases from 4 to 6 suggests that it is plausible that the zero
crossing for $B_8$ occurs close to $D=4.$
\begin{table}[bht]
\caption{Numerical values for $B_4/B_2^3$, $B_5/B_2^6$ and
$B_6/B_2^5.$ Local minima and maxima are underlined.}
\label{B456table}
\begin{tabular}{rlll}
\hline \hline $D$ & \multicolumn{1}{c}{$B_4/B_2^3$} &
\multicolumn{1}{c}{$B_5/B_2^4$} & \multicolumn{1}{c}{$B_6/B_2^5$} \\
\hline 3 &$\;\;\> 0.2869495\cdots $ &$0.110252(1)$ &$\;\;\>0.03881(6)$
\\ 4 &$ \;\;\> 0.1518460\cdots $ & $ 0.03565(5) $ &$\;\;\>0.00769(3)$
\\ 5 &$ \;\;\> 0.075978(4) $ & $ 0.01297(1) $ & $ \; \; \> 0.00094(3)
$ \\ 6 &$ \; \; \> 0.03336314\cdots $ & $ 0.007528(8) $ & $
-0.00176(2) $ \\ 7 &$ \; \; \> 0.009873(4) $ & $
\underline{0.007071(7)} $ & $ -0.00352(2) $ \\ 8 &$-0.0025576\cdots$ &
$ 0.007429(6) $ & $ -0.00451(2) $ \\ 9 &$-0.008575(3) $ & $\underline{
0.007438(6)} $ & $ \underline{-0.00478(1)} $ \\ 10 &$-0.0109624\cdots$
& $ 0.006969(5) $ & $ -0.00452(1) $ \\ 11 &$\underline{-0.011334(3)} $
& $ 0.006176(4) $ & $ -0.00395(1) $ \\ 12 &$-0.0109624\cdots$ & $
0.005244(4) $ & $ -0.003261(7) $ \\ \hline \hline
\end{tabular}
\end{table}

To proceed further we separate the Ree-Hoover diagrams with $k$ points
into classes with $m\leq k$ points which are the end points of $\tilde
f$ bonds. We designate the value of a diagram in this class, including
its combinatorial coefficient, as $B_k[m,i]$ where $i$ is an arbitrary
label specifying the graph with given $k$ and $m.$ If the graph
$B_k[m,i]$ exists for $k=m$ then it will continue to exist for $k>m.$
There is one graph in the class $B_k[0,i]$, one in the class
$B_k[4,i]$ three in the class $B_k[5,i]$, and 18 in the class
$B_k[6,i].$

For the diagram with $m=0$ the absence of $\tilde f$ bonds forces the
points to all lie within a distance $\sigma$ of each other. We define
any sequence of diagrams with increasing number of points as being
``close packed'' if the maximal volume of the convex hull of $k$
points approaches a constant as $k\rightarrow\infty$. As such, any
sequence with $m$ and $i$ fixed is a close packed sequence of
diagrams.  We have numerically studied $B_k[0,1]$, $B_k[4,1]$, and all
graphs in $B_k[5,i]$ for many values of $k$ and $D.$ Two examples are
shown in Tables \ref{cstartab} and \ref{rh2tab}.
\begin{table}[thb]
\caption[]{$B_k[0,1]/B_2^{k-1} = \wigglya/B_2^{k-1}$.}
\label{cstartab}
\footnotesize
\begin{tabular}{rllll}
\hline \hline $k$ & \multicolumn{1}{c}{$D = 2$} &
\multicolumn{1}{c}{$D = 3$} & \multicolumn{1}{c}{$D = 4$} &
\multicolumn{1}{c}{$D = 5$} \\ \hline 
4 & $0.5488(4)$ & $0.3166(3)$ &
$0.1888(2)$ & $0.1153(2)$ \\ 5 & $0.3620(3)$ & $0.1420(2)$ &
$0.0591(2)$ & $0.02522(8)$ \\ 6 & $0.2292(3)$ & $0.0593(2)$ &
$0.01648(6)$ & $0.00487(6)$ \\ 7 & $0.1412(3)$ & $0.0233(2)$ &
$0.00424(6)$ & $0.00076(1)$ \\ 8 & $0.0844(4)$ & $0.0087(2)$ &
$0.00101(2)$ & $0.000129(3)$ \\ 9 & $0.0505(4)$ & $0.00315(6)$ &
$0.000226(5)$ & $1.78(7)\times10^{-5}$ \\ 10 & $0.0293(4)$ &
$0.00111(2)$ & $5.2(2)\times10^{-5}$ & $2.5(4)\times10^{-6}$ \\ 11 &
$0.0170(3)$ & $0.000380(8)$ & $1.0(1)\times10^{-5}$ & \\ 12 &
$0.0097(2)$ & $0.000128(3)$ & $2.7(7)\times10^{-6}$ & \\ 13 &
$0.0053(1)$ & $5.2(4)\times10^{-5}$ & & \\ 14 & $0.00304(6)$ &
$1.7(3)\times10^{-5}$ & & \\ 15 & $0.00179(4)$ & & & \\
\hline \hline
\end{tabular}
\normalsize
\end{table}
\begin{table}[hbt]
\caption[]{$B_k[4,1]/B_2^{k-1}=\wigglyb/B_2^{k-1}$.}
\label{rh2tab}
\footnotesize
\begin{tabular}{rllll}
\hline \hline $k$ & \multicolumn{1}{c}{$D = 2$} &
\multicolumn{1}{c}{$D = 3$} & \multicolumn{1}{c}{$D = 4$} &
\multicolumn{1}{c}{$D = 5$} \\ \hline 4 & $-0.01644(5)$ &
$-0.02981(9)$ & $-0.0370(1)$ & $-0.0391(1)$ \\ 5 & $-0.0264(3)$ &
$-0.0316(2)$ & $-0.0270(2)$ & $-0.0189(2)$ \\ 6 & $-0.0285(5)$ &
$-0.0219(4)$ & $-0.0117(2)$ & $-0.0059(1)$ \\ 7 & $-0.0239(5)$ &
$-0.0114(2)$ & $-0.00403(8)$ & $-0.00130(3)$ \\ 8 & $-0.0183(4)$ &
$-0.0056(1)$ & $-0.00120(5)$ & \\ 9 & $-0.0123(3)$ & $-0.0025(1)$ & &
\\ 10 & $-0.0086(4)$ & & & \\ 11 & $-0.0056(2)$ & & & \\ 12 &
$-0.0038(2)$ & & & \\ 13 & $-0.0023(3)$ & & & \\ \hline \hline
\end{tabular}
\normalsize
\end{table}

In contrast to the close packed diagrams, we call diagrams $B_k[k,i]$
for which many points are forced to be more than a distance of
$\sigma$ apart ``loose packed''. One sequence of loose packed diagrams
is the $k$ point ring of $f$ bonds which we denote by ${\mathsf R}$,
for which numerical values are shown in \tab{ringtab}.  We have found
numerically for fixed $k$ that as $D\rightarrow \infty$ this ring
diagram is larger than all other studied. We therefore make the
following

{\bf Conjecture:} For fixed $k$, $\lim_{D\rightarrow\infty}
B_k/{\mathsf R}=1.$ In particular the sign of $B_k$ for large $D$ is
$(-1)^{k-1}.$

For fixed $D$, we find that for $k$ up to 7 the largest diagrams are
of the form of a ring with ``insertions'' where one point of the ring
is replaced by a cluster of points connected by $f$ and $\tilde{f}$
bonds. We thus may define a sequence of loose packed diagrams which
consist of ring diagrams with an inserted diagram. Note that the
contribution of these sequences to the virial coefficient alternate in
sign, whereas close packed diagrams always contribute with the same
sign. The smallest possible insertions have 4 points, of which there
are two examples. The larger of these is given in \tab{ivbtab} where
the type of insertion is indicated in the figure caption. Multiple
insertions are possible but never dominate for the values of
$k$ we have studied.
\begin{table}[bht]
\caption[]{$B_k[k,1]/B_2^{k-1}={\mathsf R}/B_2^{k-1}$. The underline
marks the approximate location of the minimum value.}
\label{ringtab}
\footnotesize
\begin{tabular}{rllll}
\hline \hline $k$ & \multicolumn{1}{c}{$D = 2$} &
\multicolumn{1}{c}{$D = 3$} & \multicolumn{1}{c}{$D = 4$} &
\multicolumn{1}{c}{$D = 5$} \\ \hline 4 & $-0.01639(9)$ & $-0.0298(1)$
& $-0.0371(1)$ & $-0.03925(5)$ \\ 5 & $\;\;\>0.00860(6)$ &
$\;\;\>0.01623(9)$ & $\;\;\>0.0214(2)$ & $\;\;\>0.0230(1)$ \\ 6 &
$-0.00526(8)$ & $-0.0109(1)$ & $-0.0150(2)$ & $-0.01689(9)$ \\ 7 &
$\;\;\>0.00335(6)$ & $\;\;\>0.0078(2)$ & $\;\;\>0.0124(2)$ &
$\;\;\>0.0142(3)$ \\ 8 & $-0.00234(5)$ & $-0.0064(1)$ & $-0.0106(2)$ &
$-0.0129(3)$ \\ 9 & $\;\;\>0.00177(4)$ & $\;\;\>0.0053(1)$ &
$\;\;\>0.0098(2)$ & $\;\;\>0.0126(2)$ \\ 10 & $-0.00125(3)$ &
$-0.00452(9)$ & $-0.0091(2)$ & \underline{$-0.0126(3)$} \\ 11 &
$\;\;\>0.00095(2)$ & $\;\;\>0.00392(8)$ & $\;\;\>0.0089(2)$ &
$\;\;\>0.0128(3)$ \\ 12 & $-0.00074(1)$ & $-0.00333(7)$ &
\underline{$-0.0083(2)$} & $-0.0134(3)$ \\ 13 & $\;\;\>0.00055(1)$ &
$\;\;\>0.00313(8)$ & $\;\;\>0.0086(2)$ & $\;\;\>0.0142(3)$ \\ 14 &
$-0.00041(1)$ & $-0.0027(1)$ & $-0.0086(2)$ & $-0.0166(3)$ \\ 15 &
$\;\;\>0.00033(2)$ & \underline{$\;\;\>0.0026(1)$} & $\;\;\>0.0087(3)$
& $\;\;\>0.0183(4)$ \\ \hline \hline
\end{tabular}
\normalsize
\end{table}
\begin{table}[htb]
\caption[]{$B_k[k,2]/B_2^{k-1}={{\mathsf R}
\left(\insivb\right)}/B_2^{k-1}$.  The underline marks the approximate
location of the minimum value.}
\label{ivbtab}
\footnotesize
\begin{tabular}{rllll}
\hline\hline $k$ & \multicolumn{1}{c}{$D = 2$} & \multicolumn{1}{c}{$D
= 3$} & \multicolumn{1}{c}{$D = 4$} & \multicolumn{1}{c}{$D = 5$} \\
\hline 5 & $-0.01017(4)$ & $-0.01654(7)$ & $-0.01748(6)$ &
$-0.01551(4)$ \\ 6 & $\;\;\>0.00756(6)$ & $\;\;\>0.01191(8)$ &
$\;\;\>0.01257(6)$ & $\;\;\>0.01067(6)$ \\ 7 & $-0.00607(8)$ &
$-0.01003(7)$ & $-0.0106(1)$ & $-0.00936(9)$ \\ 8 & $\;\;\>0.0051(1)$
& $\;\;\>0.0088(2)$ & \underline{$\;\;\>0.0100(1)$} &
\underline{$\;\;\>0.00902(8)$} \\ 9 & $-0.00427(8)$ & $-0.0083(1)$ &
$-0.0100(1)$ & $-0.0095(2)$ \\ 10 & $\;\;\>0.00381(8)$ &
$\;\;\>0.0082(2)$ & $\;\;\>0.0106(2)$ & $\;\;\>0.0106(2)$ \\ 11 &
$-0.00309(3)$ & $-0.0075(1)$ & $-0.0109(2)$ & $-0.0116(2)$ \\ 12 &
$\;\;\>0.00259(3)$ & $\;\;\>0.00728(7)$ & $\;\;\>0.0115(2)$ &
$\;\;\>0.0136(3)$ \\ 13 & $-0.00220(3)$ & $-0.0071(3)$ & $-0.0125(3)$
& $-0.0156(3)$ \\ 14 & $\;\;\>0.00188(4)$ & $\;\;\>0.0068(4)$ &
$\;\;\>0.0140(5)$ & $\;\;\>0.0172(4)$ \\ 15 & $-0.00151(5)$ &
$-0.0060(5)$ & $-0.0145(7)$ & $-0.0227(7)$ \\ 
\hline \hline
\end{tabular}
\normalsize
\end{table}

The data of Tables \ref{cstartab}--\ref{ivbtab} differ in the two
cases $D=2$ and $D\geq 3$ in the following respect. For $D=2$ the
close packed diagrams $B_k[0,1]$ and $B_k[4,1]$ are larger than the
loose packed diagrams $B_k[k,i]$ we have considered, whereas for
$D\geq 3$ the loose packed diagrams rapidly dominate the close packed
diagrams for large $k$. Consequently we will consider the two cases
separately.

For $D=2$ an examination of Tables \ref{cstartab}--\ref{ivbtab}
indicates that all diagrams are roughly of the same order of
magnitude. A measure of the relative size of the close packed to the
loose packed diagrams is given by the ratio $|B_k[k,1]/B_k[0,1]|$
which increases from $0.0299$ to $0.184$ as $k$ increases from 4 to 15
($0.382$ when $k=17$). It can be argued that if this increase
continues then eventually the loose packed diagrams will dominate, but
such a conclusion requires further computations. The data can be
interpreted to say there is a qualitative difference between $D=2$ and
$D\geq 3.$ This interpretation may be correct but is not compelling.

For $D\geq 3$ the loose packed diagrams $B_k[k,i]$ rapidly dominate
for large $k$ the close packed diagrams $B_k[m,i]$ for fixed $m$ and
in particular $|B_k[k,1]|$ first becomes larger than $B_k[0,1]$ for
$k=9$ in $D=3$, $k=7$ in $D=4$, and $k=6$ in $D=5$.  Further studies
indicate that $B_k[k,i]$ even dominates $B_k[k-1,i]$ for large $k.$
Thus in order to determine the radius of convergence we will restrict
our attention to the diagrams belonging to $B_k[k,i].$

We see in \tab{ringtab} that the Ree-Hoover ring diagram
$B_k[k,1]/B_2^{k-1}$ increase in magnitude for $D\geq 4$ when $k$ is
sufficiently large. In $D=3$ the data of \tab{ringtab} can be
interpreted as either having a minimum near $k=1$ or possibly
approaching a constant as $k\rightarrow \infty$.

The largest diagram shown is the four point insertion $B_k[k,2]$ of
\tab{ivbtab}. It seems clear that for $D\geq 4$ this diagram will
always be bigger than the Ree-Hoover ring but the ratio seems to grow
only as a power of $k$.  We believe the same behavior happens for
$D=3$ but it is more difficult to see with the precision given in the
tables.

From these and other numerical studies we conjecture that the
exponential rate of growth of the ring and ring diagrams with
insertions is the same.

To estimate the rate of growth we concentrate on the Ree-Hoover ring
of \tab{ringtab}. We note that the absolute value of this diagram must
be strictly less than the absolute value of the Mayer ring diagram
which is obtained by replacing all the $\tilde f$ bonds by unity.  The
Mayer ring may be expressed as a single integral~\cite{katsura1963a},
and from this we obtain \be
\left|\frac{B_k[k,1]}{B_2^{k-1}}\right|\leq{\frac{(k-1)(2\pi)^{\frac{kD}{2}}}{
2k\Omega_{D-1}^{k-2}}} \int_0^\infty \! \!\!\! dx
x^{D-1}\left[\frac{J_{D/2}(x)}{x^{D/2}}\right]^{k}, \ee where
$\Omega_{D-1}\equiv{2\pi^{D/2} / \Gamma(D/2)}$ and $J_{D/2}(k)$ is the
Bessel function of the first kind. The large $k$ behavior of this
integral is obtained by steepest descents and thus we have as
$k\rightarrow \infty$ \be
\left|\frac{B_k[k,1]}{B_2^{k-1}}\right|\leq{\frac{(k-1)(1+D/2)^{D/2}}{
k^{1+D/2}\Gamma(1+D/2)}} \> 2^{k-2}.  \ee

From \tab{ringtab} we see that for all $D$ the ratios
$B_{k+1}[k+1,1]/(B_2B_k[k,1])$ are all substantially below this bound
of $2$. If the data in \tab{ringtab} are extrapolated to say for $D=3$
that $B_k[k,1]$ goes to a constant (or a power of $k$) as $k
\rightarrow \infty$ then we see that the radius of convergence of the
sum of Ree-Hoover ring diagrams is $\eta_{rh}=0.25.$ At worst the
ratios in \tab{ringtab} for $D=3$ are bounded below by $0.91$ as
$k\rightarrow \infty$ which leads to a radius of convergence of
$\eta_{rh}=0.27$.  This radius of convergence is substantially less
than the freezing density of $\eta_f=0.49.$ Similarly we estimate from
\tab{ringtab} that in $D=4$ the radius of convergence of Ree-Hoover
rings is $\eta_{rh}\sim0.12$ and in $D=5$ is $\eta_{rh}\sim0.052$
which are to be compared with the freezing densities $\eta_f=0.31$ in
$D=4$ and $\eta_f=0.19$ in $D=5$ obtained
from~\cite{michels1984a,luban1990a,finken2001a}.  The alternation of
sign of the ring means that the leading singularity
of the sum of these diagrams is on the negative $\eta$ axis.

If all loose packed diagrams $B_k[k,i]$ had the sign $(-1)^{k-1}$ of
the Ree-Hoover ring diagram then the
estimates of $\eta_{rh}$ found above would be an upper bound on the
radius of convergence of the virial expansion. However for every order
$k$ both signs occur in the class $B_k[k,i]$ and in particular the
diagram $B_k[k,2]$ is larger than and has the opposite sign from
$B_k[k,1].$ Therefore cancellations of diagrams within the class
$B_k[k,i]$ can occur and as evidence of such cancellation we note that
the virial coefficient $B_8/B_2^7$ of \tab{virialtable} for $D=3$ is
smaller than both $B_8[8,1]$ and $B_8[8,2]$, and has the opposite sign
to the ring diagram $B_8[8,1]$.

However, the diagrams $B_k[k,2]$ also alternate in sign, and by
themselves lead to a singularity on the negative real axis.  If this
singularity is to be avoided extensive cancellation must take place
beyond what can be seen in the virial coefficients up through order 8
as given in \tab{virialtable}. From this point of view we see that a
detailed study of diagrams for orders substantially greater than 8 is
needed to substantiate any claim concerning the radius of convergence
of the virial series for hard spheres.  None of the approximate
equations of state for hard
spheres~\cite{thiele1963a,wertheim1963a,guggenheim1965a,carnahan1969a,hoover1968a,ree1964a,vanrensburg1993a,hoste1984a,goldman1988a,lefevre1972a,ma1986a,jasty1987a,song1988a}
includes diagrams of such high order. We therefore conclude that
there is no existing evidence to support the claim that the virial
expansion converges beyond the freezing density $\eta_f$ of hard
spheres.

\acknowledgments{This work was supported in part by the National
Science Foundation under DMR-0073058. We thank Prof.~R.~J.~Baxter and
Prof.~G.~Stell for useful discussions.}


\end{document}